\documentclass[aps,prd,nofootinbib,preprint,floatfix,superscriptaddress]{revtex4-1}
\usepackage{setspace}
\usepackage{graphicx}
\usepackage{dcolumn}
\usepackage{multirow}
\usepackage{subfigure}
\usepackage{times,mathptm}
\usepackage{float}
\usepackage{color}
\usepackage{amsmath,amsfonts}
\usepackage{mathptmx}
\usepackage{mathrsfs}
\usepackage{bbm}
\usepackage{bm}
\usepackage{xfrac}

\newcommand{\bea}{\begin{eqnarray}}
\newcommand{\eea}{\end{eqnarray}}

\renewcommand{\d}{\delta}
\newcommand{\F}{{\cal F}}

\renewcommand{\l}{\lambda}

\renewcommand{\b}{\beta}
\renewcommand{\a}{\alpha}

\renewcommand{\o}{\omega}
\newcommand{\tr}{\text{Tr}}

\newcommand{\vx}{{\vec{x}}}
\newcommand{\vy}{{\vec{y}}}

\newcommand{\vk}{{\vec{k}}}

\newcommand{\vn}{\vec{n}}
\newcommand{\m}{\mu}
\newcommand{\g}{\gamma}

\newcommand{\s}{\sigma}
\renewcommand{\k}{\kappa}

\renewcommand{\vn}{\vec{n}}

\newcommand{\N}{{\cal N}}

\newcommand{\oh}{\frac{1}{2}}
\newcommand{\oq}{\frac{1}{4}}

\newcommand{\dg}{\dagger}
\newcommand{\non}{\nonumber}

\newcommand{\rf}[1]{(\ref{#1})}
\newcommand{\ra}{\rightarrow}
\newcommand{\pa}{\partial}
\renewcommand{\vec}[1]{\bm #1}
\usepackage{ulem}

\begin{document}

\title{Numerical study of the Yang-Mills vacuum wavefunctional \\ in D=3+1  dimensions} 
 
\author{Jeff Greensite}
\affiliation{Physics and Astronomy Dept.,
San Francisco State University, San Francisco CA 94132, USA}
\author{{\v S}tefan Olejn\'{\i}k}
\affiliation{Institute of Physics, Slovak Academy of Sciences, SK--845 11, Bratislava, Slovakia}

\date{\today}
\vspace{60pt}
\begin{abstract}

\singlespacing
 
   Ratios of the true Yang-Mills vacuum wavefunctional, evaluated on any two field configurations out of a finite set of configurations, can be obtained from lattice Monte Carlo simulations.  The method was applied some years ago to test various proposals for the vacuum wavefunctional in 2+1 dimensions.  In this article we use the same method to test our own proposal for the Yang-Mills ground state in 3+1 dimensions.  This state has the property of ``dimensional reduction" at large scales, meaning that the (squared) vacuum state, evaluated on long-wavelength, large scale fluctuations, has the form of the Boltzmann weight for Yang-Mills theory in $D=3$ Euclidean dimensions.  Our numerical results support this conjectured behavior.  We also investigate the form of the ground state evaluated on shorter wavelength configurations.

\end{abstract}

\maketitle

\singlespacing
%\begin{widetext}
\section{\label{sec:intro}Introduction}

   Many years ago it was suggested \cite{Greensite:1979yn} that long-wavelength vacuum fluctuations in $D=3+1$ Yang-Mills theory might be controlled, in temporal gauge, by a vacuum wavefunctional of the form
\bea
    \Psi_0[A] = {\cal{N}} \exp\left[ - \oh \m \int d^3x ~ \tr F_{ij}^2(x) \right]
\label{DRc}
\eea
where $\m$ is a constant with dimensions of inverse mass, ${\cal{N}}$ is a normalization constant, and 
$F_{ij}=\pa_i A_j - \pa_j A_i - i[A_i,A_j]$.  This idea is known as ``dimensional reduction,"  since the vacuum expectation value of an operator $Q$ on a time slice $\langle \Psi_0 |Q| \Psi_0 \rangle$ is clearly the same as the expectation value of the operator in Yang-Mills theory in $D=3$ Euclidean dimensions.\footnote{A similar proposal was made by Halpern \cite{Halpern:1978ik} in $D=2+1$ dimensions.}  The idea was tested numerically on rather small lattices \cite{Greensite:1988rr},  with results which appeared to support the suggestion.  

    However, a vacuum wavefunctional of the form \rf{DRc} is obviously not correct for small scale, high-frequency fluctuations, where we may expect asymptotic freedom to come into play.  For a free abelian theory, the ground state is well known, and is quite different from the dimensional reduction form:
\bea
 \Psi_0[A] = {\cal{N}} \exp\left[ - {1\over 4e^2} \int d^3x d^3y ~ F_{ij}(x) \left({1\over \sqrt{-\nabla^2}} \right)_{xy} F_{ij}(y) \right]
\label{abelian}
\eea
It is natural to guess that the true Yang-Mills vacuum wavefunctional  in temporal gauge might have a structure which in some way interpolates between these two forms.  In ref.\ \cite{Greensite:2007ij} we proposed that in 2+1 dimensions
\bea
 \Psi_0[A] = {\cal{N}} \exp\left[ - {1\over 4g^2} \int d^2x d^2y ~ F^a_{ij}(x) \left({1\over \sqrt{-D^2 - \l_0 + m^2}} \right)^{ab}_{xy} F^b_{ij}(y) \right]
\label{GO}
\eea
might be a reasonable approximation to the true vacuum wavefunctional, where $a,b$ are color indices, $D^2$ is the covariant Laplacian in the adjoint representation, $\l_0$ is the lowest eigenvalue of $-D^2$ in a given configuration, and $m$ is a parameter with dimensions of mass.  The physical state condition in temporal gauge requires gauge invariance of all physical wavefunctionals (at least with respect to infinitesimal gauge transformations), a property which 
is evident in  \rf{GO}.  The same proposal, but without the $\l_0$ subtraction, was made by Samuel \cite{Samuel:1996bt}. The motivation for the $\l_0$ subtraction is that $-D^2$ has a positive definite spectrum, finite with a lattice regularization, with the lowest eigenvalue tending to infinity for typical configurations in the continuum limit.  Thus, a non-zero kernel in the continuum limit requires a subtraction of this kind, otherwise $\Psi_0$ would tend to the infinite strong-coupling vacuum in the continuum limit.

    This proposal for the ground state was tested numerically a few years ago, by a method which will be explained in the next section, and the results were encouraging \cite{Greensite:2011pj}.\footnote{Another proposal in 2+1 dimensions is due to Karabali, Kim, and Nair \cite{Karabali:1998yq}.  Their form of the vacuum state is not gauge-invariant, at least as originally proposed.  See ref.\  \cite{Greensite:2011pj} for a further discussion.}  In this article we will use the same techniques to study the naive extension of the state \rf{GO} to $3+1$ dimensions, and further test the hypothesis of dimensional reduction.

\section{The relative weights method}

    We work in a lattice regularization.  The squared vacuum state can be expressed in the path integral form
\bea
\Psi_0^2[U'] = {1\over Z} \int DU  \prod_{x} \prod_{i=1}^3 \d[U_i(x,0)-U'_i(x)] e^{-S}
\label{Up}
\eea
Although the direct numerical evaluation of the path integral in \rf{Up} is difficult, the numerical calculation of a ratio of
$\Psi_0^2[U'']/\Psi_0^2[U']$ (the ``relative weight'') is actually straightforward, assuming that configurations $U''$ and $U'$ are nearby in configuration space, so that the relative weight (or its inverse) is not too small.  Consider a set of $M$ such configurations 
\bea
{\cal U} = \{ U_i^{(j)}(x), j=1,2,...M \}
\label{theset}
\eea
Each member of the set can be used to specify the spacelike links on the timeslice $t=0$.  Let us now make the rescaling
\bea
        \widetilde{\Psi}_0^2[U^{(n)}] &=& {\Psi_0^2[U^{(n)}] \over \sum_{m=1}^M \Psi_0^2[U^{(m)}] }
\non \\
&=& {\int DU  \prod_{x} \prod_{i=1}^3 \d[U_i(x,0)-U^{(n)}_i(x)] e^{-S} \over
\sum_{m=1}^M \int DU  \prod_{x} \prod_{i=1}^3 \d[U_i(x,0)-U^{(m)}_i(x)] e^{-S} }
\eea
This is a statistical system, with the configurations on the $t=0$ timeslice restricted to the finite set ${\cal U}$, and
$\widetilde{\Psi}_0^2[U^{(n)}]$ has the interpretation of a probability that the $n$-th configuration will appear in the timeslice.   The system can be simulated numerically, using the usual heatbath for spacelike links at $t\ne 0$, and for timelike links, while the spacelike links at $t=0$ are updated simultaneously, selecting one of the set of $M$ configurations \rf{theset} at random and accepting or rejecting according to the Metropolis algorithm.  To get a reasonable acceptance rate, it is necessary that the configurations in the set \rf{theset} are nearby in lattice configuration space.  If we let $N_n$ denote the number of times the $n$-th configuration is accepted, with $N_{tot}=\sum_{m=1}^M N_m$ the total number of updates, then
\bea
 \widetilde{\Psi}_0^2[U^{(n)}]  = \lim_{N_{tot} \ra \infty} {N_n \over N_{tot}}
\eea
Since $ \widetilde{\Psi}_0$ is simply a rescaling of $\Psi_0$, the corresponding relative weights are also
\bea
 {\Psi_0^2[U^{(m)}] \over \Psi_0^2[U^{(n)}]}   = \lim_{N_{tot} \ra \infty} {N_m \over N_n}
\label{relw}
\eea

   The relative weights method outlined above was originally proposed in ref.\ \cite{Greensite:1988rr}.  
Using this method, we can test any proposal for the vacuum state, of the form
\bea
          \Psi^2_0[U] = {\cal N} e^{-R[U]}
\eea
by plotting
\bea
               -\log\left[{N_n \over N_{tot}}\right]  ~~~~ \mbox{vs.}~~~~ R[U^{(n)}] 
\eea
If the proposal is correct, the data should fall on a straight line with unit slope.  
\section{Results}

    We specialize to the SU(2) gauge group.  Taking the lattice-regularized field strength to be
\bea
          F^a_{ij}(x) = -i \tr[U_i(x) U_j(x+\hat{i})U^\dg_i(x+\hat{j})U^\dg_j(x) \s^a]
\label{F}
\eea
our proposal for the Yang-Mills vacuum wavefunctional on the lattice, in 3+1 dimensions, is
\bea
\Psi_0[U]  = {\cal N} \exp\left[ - {c\over 8} \sum_x \sum_y \sum_{i<j}  F_{ij}^a(x) \left({1\over \sqrt{-D^2 - \l_0 + m^2}} \right)^{ab}_{xy} F^b_{ij}(y) \right]
\label{PsiU}
\eea
where $D^2$ is the lattice-regularized covariant Laplacian
\bea
         D^2_{xy} = \sum_{\m} \Bigl[ U_\m(x) \d_{y,x+\hat{\m}}
         + U^{\dg}_\m(x-\hat{\m}) \d_{y,x-\hat{\m}}  - 2 \mathbbm{1} \d_{xy} \Bigl]
\eea
in the adjoint representation.
In 2+1 dimensions we have identified $c=\b=4/g^2$, which scales as the inverse lattice spacing at weak couplings. In 3+1 dimensions, however, we just take $c$ to be a parameter which depends on the lattice spacing in a manner to be determined.   

\subsection{Non-abelian constant configurations}

\begin{figure}[t!]
\includegraphics[scale=0.8]{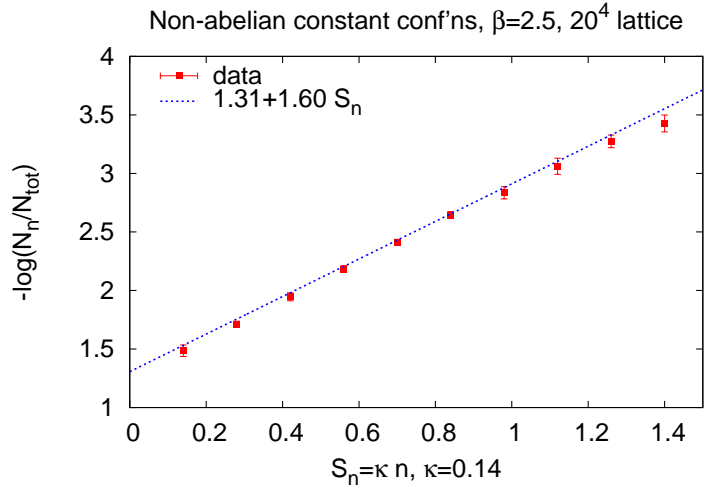}
\caption{A plot of $\log(N_n/N_{tot})$ vs.\ the sum over plaquette terms $S_n$, for non-abelian constant configurations.
The slope of the straight line fit through the data points determines the coefficient $\mu$ in $R[U]=\m S$.}
\label{calc_mu}
\end{figure}

   To apply the relative weights method, we begin by choosing a set of non-abelian constant configurations, for which the 
$U_i(x)$ are constant in space, but $[U_i,U_j] \ne 0$ for $i \ne j$.  The set is
\bea
{\cal U}_{nac} = \Bigl\{ U_k^{(n)}(x)= \sqrt{1- (a^{(n)})^2}  \mathbbm{1} + i a^{(n)} \s_k , 
                          ~~a^{(n)} = \left({n\k \over 6L^3}\right)^{1/4}, ~~ n=1,2,...M \Bigr\}
\eea
For small amplitude configurations (i.e.\ $\k$ sufficiently small), and taking $\Psi_0^2 = \N \exp[-R[U]]$, eq.\ \rf{PsiU}  gives us
\bea
R[U] = {c \over 4m} \sum_x \sum_{i<j} (F^a_{ij})^2
\label{R1}
\eea
From \rf{F}
\bea
U_i(x) U_j(x+\hat{i})U^\dg_i(x+\hat{j})U^\dg_j(x) = \sqrt{1 - \oq (F^a_{ij})^2} \mathbbm{1} + i F^a_{ij} {\s^a \over 2}
\eea
and therefore
\bea
\oh \tr[U_i(x) U_j(x+\hat{i})U^\dg_i(x+\hat{j})U^\dg_j(x)] &=& 1 - {1\over 8} (F^a_{ij})^2 + O[(F^a_{ij})^4]
\eea
Disregarding the $O(F^4)$ term, we have
\bea
R[U] = {2c \over m} \sum_{plaq}\Bigl(1- \oh \tr[U_i(x) U_j(x+\hat{i})U^\dg_i(x+\hat{j})U^\dg_j(x)] \Bigr)
\label{R2}
\eea
which implies the dimensional reduction form
\bea
\Psi^2_0[U] = \N \exp\left[- \m \sum_{plaq} \left( 1 -\oh \tr[U_i(x) U_j(x+\hat{i})U^\dg_i(x+\hat{j})U^\dg_j(x)]\right) \right]
\label{DR}
\eea
with $\m=2c/m$.  We can determine $\m$ at any given $\b$ by plotting 
\bea
-\log\left[{N_n \over N_{tot}}\right]  ~~~~ \mbox{vs.}~~~~
        L^3 \sum_{i<j} \left(1-\oh\tr[U^{(n)}_i U^{(n)}_j U^{(n) \dg}_i U^{(n)\dg}_j] \right)
\eea
and identifying $\mu$ with the slope of the best straight line fit through the data points, as shown in Fig.\ \ref{calc_mu}.
For the non-abelian constant configurations shown above
\bea
        S_n \equiv L^3 \sum_{i<j} (1 - \oh\tr[U^{(n)}_i U^{(n)}_j U^{(n) \dg}_i U^{(n)\dg}_j]) =  \kappa n
\eea

\begin{figure}[t!]
\includegraphics[scale=0.8]{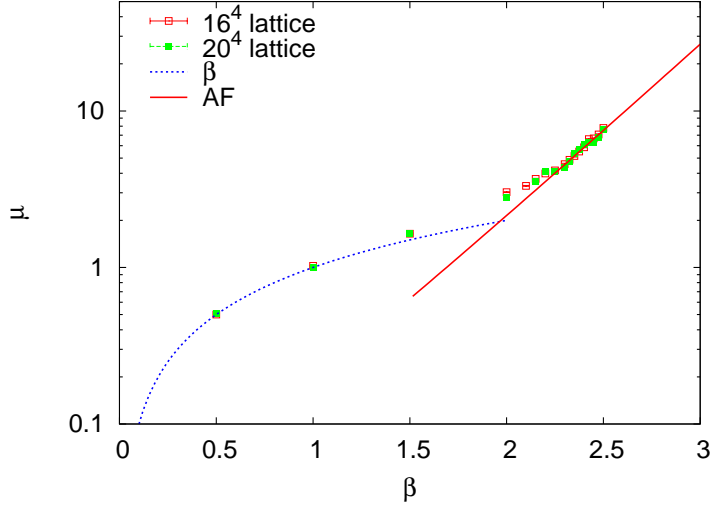}
\caption{Variation of $\m$ with $\b$.  The dotted line is the strong-coupling prediction, while the straight line fit, given by
eqs.\  \rf{mb} and \rf{fb}, is the asymptotic freedom prediction.}
\label{mvb}
\end{figure}  

     Figure \ref{mvb} is a plot of $\mu$ vs.\ $\b$, with $\m$ determined by the relative weights method just described.
Since $\m = \mu_{phys}/a$, where $a$ is the lattice spacing, we expect that at weak couplings
\bea
          \m(\b) = \m_0 f^{-1}(\b)
\label{mb}
\eea
where
\bea
                    f(\b) = ({6\over 11} \pi^2 \beta)^{51/121} \exp(-{3\over 11} \pi^2 \beta)
\label{fb}
\eea
and this appears to be entirely consistent with our weak coupling data, with $\m_0 \approx 0.0269$.   This is just an improvement, with larger lattices and better statistics, of the dimensional reduction test reported long ago in 
ref.\ \cite{Greensite:1988rr}.  However, in terms of our improved wavefuctional  \rf{PsiU}, we must now make the identification
$\m = 2 c/m$, and see if this identification is consistent with the constants $c$ and $m$ obtained from other sets of configurations, going beyond the dimensional reduction limit.

\subsection{Abelian plane wave configurations}

     We now consider abelian plane wave configurations of the form
\bea
{\cal U}_{apw} &=& \Bigl\{ U_1^{(m)}(x)= \sqrt{1- \left(a^{(m)}_{\vn}(x)\right)^2}  \mathbbm{1} + i a^{(m)}_{\vn}(x) \s_3, 
~~~ U_2^{(m)}(x)=U_3^{(m)}(x) = \mathbbm{1} \Bigr\}
\non \\
        a^{(m)}_{\vn}(x) &=& L^{-3/2} \sqrt{\a + \g m }\cos(2\pi \vx \cdot \vn/L)
\label{apw}
\eea
with $\vn = (n_1,n_2,n_3)$ the mode numbers, and $m=1,2,..,10$.  In this case $\Psi_0^2 = \N \exp[-R[U]]$ with
\bea
        R[U] &=& {c\over 4} \sum_x \sum_y \sum_{i<j} F^a_{ij}(x) G^{ab}_{xy} F^b_{ij}(y)  
\non \\
        G^{ab}_{xy} &=& \sum_{q} {1 \over \sqrt{ \l_q - \l_0 + m^2}} \phi_q^a(x) \phi_q^b(x)    
\label{R0}
\eea
where
\bea
              (-D^2)^{ab}_{xy} \phi_q^b(y) = \l_q \phi^a_q(x)
\eea
is the eigenvalue equation for the lattice Laplacian operator, with $\l_0$ the smallest eigenvalue.  For the case of abelian configurations oriented in, say, the color 3 direction, which is true for the abelian plane wave configurations 
\rf{apw}, there is a set of solutions
\bea
           \phi^a_k(x) = \sqrt{1\over L^3} \d^{a3} e^{ik\cdot x}, ~~~  \l_k  =  k_L^2, ~~~ \vec{k} = {2\pi \over L} \vec{n}
\eea
where
\bea
         k_L = \sqrt{4 \sum_{i=1}^3 \sin^2(\oh k_i)}
\eea
is the lattice momentum.  This set is not all of the eigenstates, but only one third of them.  However, these eigenstates are all pointing in the 3-direction of color space, and it is not hard to see that orthogonality implies that every other eigenstate must point in the 1-2 color plane.  Since in our case the $F_{ij}^a(x)=-i\tr[U_{ij}(x) \s_a]$ are proportional to $\d^{a3}$, only the eigenstates with non-zero components in the color-3 direction contribute to $R[U]$, which is the set of eigenstates shown.     
For these eigenstates $G^{ab}_{xy}=\d^{a3} \d^{b3}G(x-y)$, where        
\bea
         G(x-y) &=& {1\over L^3} \sum_k {1  \over \sqrt{k_L^2 + m^2}} e^{i k \cdot (x-y)}
\label{G0}
\eea
It turns out that a good fit to the data will actually require a slight generalization, and therefore a modification of the ansatz
\rf{PsiU}.  We will take
\bea
G^{ab}_{xy} &=& \sum_{q} {1 + d \sqrt{\l_q - \l_0} \over \sqrt{ \l_q - \l_0 + m^2}} \phi_q^a(x) \phi_q^b(x)
\eea
which, for the abelian plane wave configurations, reduces to $G^{ab}_{xy}=\d^{a3} \d^{b3} G(x-y)$, where
\bea
                  G(x-y) &=& {1\over L^3} \sum_k {1 + d k_L \over \sqrt{k_L^2 + m^2}} e^{i k \cdot (x-y)}
\label{Gd}
\eea
With this generalization, we have for the set \rf{apw}
\bea
         R[U^{(n)}] = \oh (\a + \g n) k_L^2 {c (1+d k_L )\over \sqrt{k_L^2 + m^2}}
\eea
We now plot
\bea
             -\log{N_m \over N_{tot}} ~~~\mbox{vs.}~~~  \oh (\a + \g m)
\eea
and again fit a straight line through the data.  Denote the slope by $\o(k_L)$.  Then we want to see whether, at each $\b$,
the data for $\o(k_L)$ can be fit by
\bea
            \o(k_L) = k_L^2 {c (1+d k_L )\over \sqrt{k_L^2 + m^2}}
\label{om1}
\eea
If so, then the data for $-\log(N_m/N_{tot})$ vs.\ $R[U^{(m)}]$ has unit slope, as required.  We then
study the $\b$ dependence of $c, d, m$.  

\begin{figure}[t!]
\subfigure[~$\b=2.4$]  % caption for subfigure a
{   
 \label{b24}
 \includegraphics[scale=0.6]{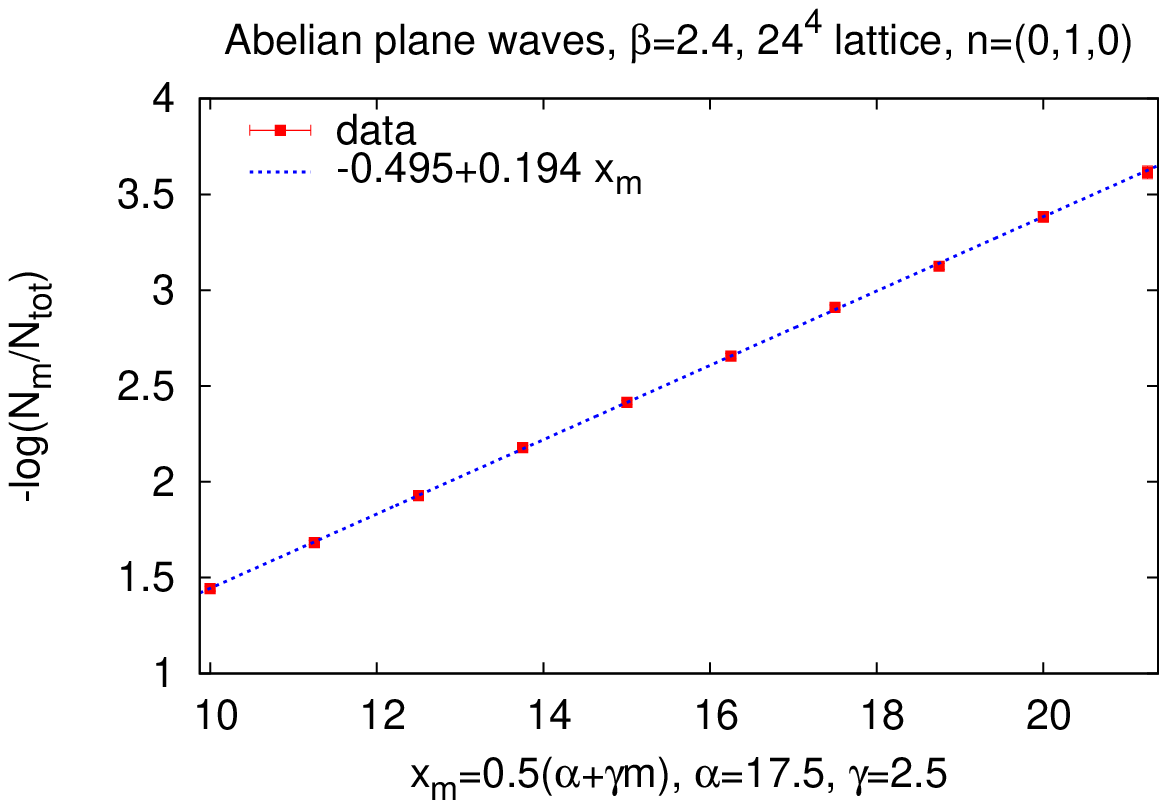}
}
\subfigure[~$\b=2.5$]  % caption for subfigure a
{   
 \label{b25}
 \includegraphics[scale=0.6]{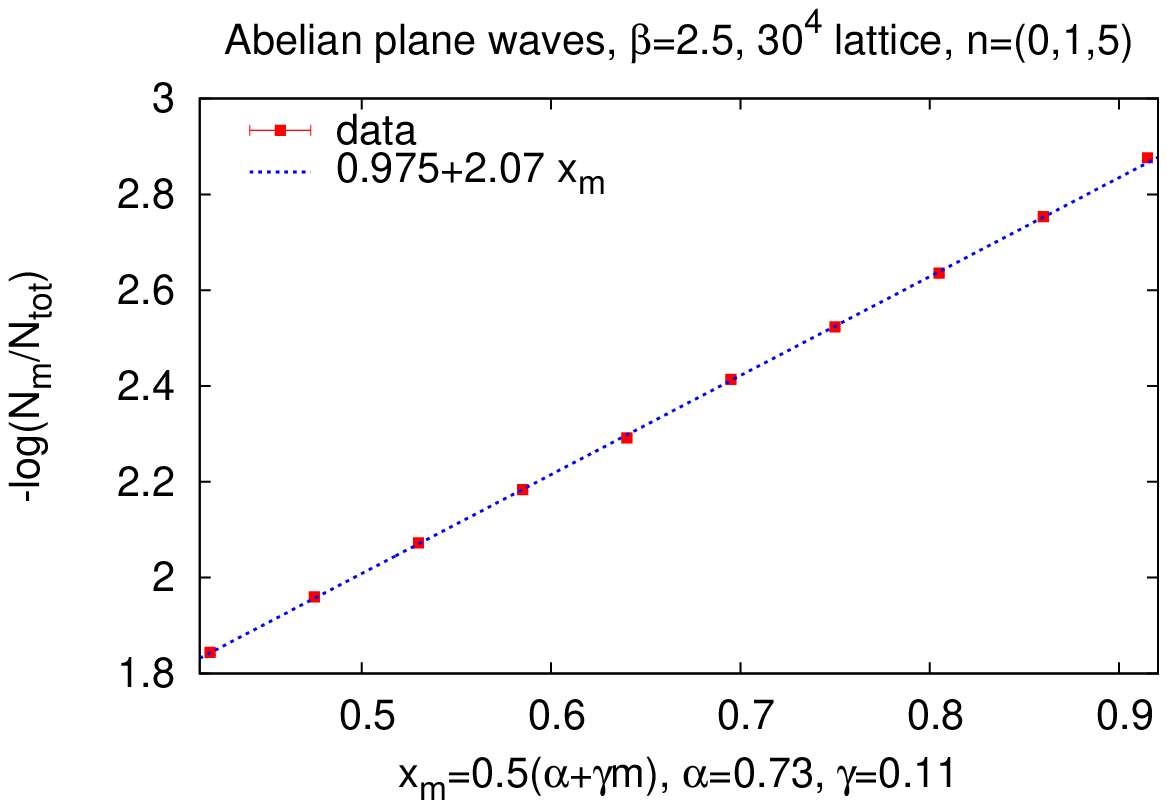}
}
\caption{The slope of the straight line fit to $-\log(N_m/N_{tot})$ vs.\ $\oh (\a + \g m)$ determines $\o(k_L)$ at a given
$\vec{k}$ and $\b$.  Here we display the examples of the data for (a) $\b=2.4$, and (b) $\b=2.5$, with momenta 
$\vec{k} = 2\pi \vec{n}/L$ corresponding to mode numbers ${\bf n}=(0,1,0)$ and (0,1,5) respectively. }
\label{sample}
\end{figure}

\begin{figure}[h!]
\includegraphics[scale=0.8]{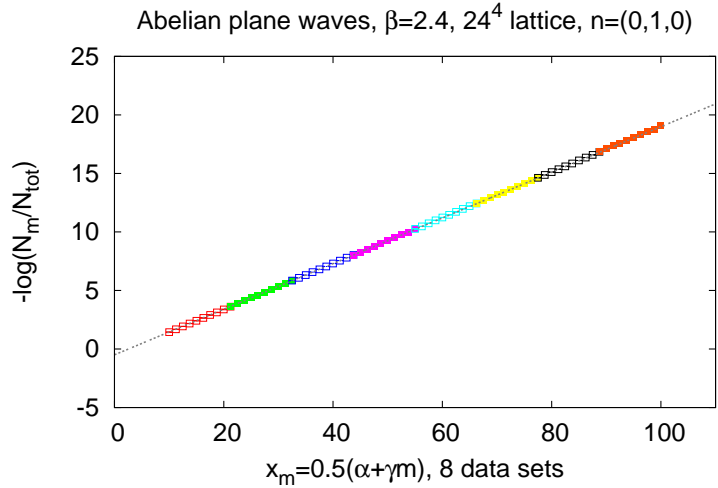}
\caption{The slope determined from $-\log(N_m/N_{tot})$ does not strongly depend on the choice of parameters $\a$
and $\g$ for abelian plane waves. Eight sets of configurations (in various colors) are shown here, each at the same value of 
$\b=2.4$ and wavevector corresponding to ${\mathbf n}=(0,1,0)$, but with different choices of $\a,~\g$.  Note the range of the $y$-axis.  An overall constant is added to the values of $-\log(N_m/N_{tot})$ in each set, such that the value for the last configuration in one set coincides with that of the first configuration in the next set. The straight-line fit shown in the figure
comes from the first data set (red open squares). The variation of slopes obtained from each separate data set is very small, on the order of 2\%.}
\label{linearity}
\end{figure}

    Fig.\ \ref{sample} shows sample plots of $-\log(N_m/N_{tot})$ vs.\ 
$\oh (\a + \g m)$;  the corresponding $\o(k_L)$ is given by the slope of the straight-line fit. 
We choose the range of ${\oh (\a + \g m)}$ so that the variation in $N_m/N_{tot}$ is not too large,  i.e.\ an order of magnitude or so, and in general the range of {$\oh (\a + \g m)$} needed to fulfill this condition will depend on the mode numbers. One might worry that the linear fit to the $-\log(N_m/N_{tot})$ vs.\ $\oh (\a + \g m)$ data might only work in a narrow window, and that we are really only looking at the tangent to a curve, whose slope might be different for different choices of $\a, \g$.  This does not appear to be a problem.  We have verified that these data sets are in fact linear for variation in $N_m/N_{tot}$ over many orders of magnitude.  This is seen in Fig.\ \ref{linearity}, where we have juxtaposed the data for eight data sets at $\b=2.4$.  Each data set is for configurations with ${\mathbf n}=(0,1,0)$, but each corresponds to taking a different range of ${\oh (\a + \g m)}$. 
In the plot, the variation in $N_m/N_{tot}$ runs over seven orders of magnitude.  The data is chosen 
such that the $\oh (\a + \g m)$ value of the last configuration of one set coincides with the corresponding value of the first configuration of the next set.  The data sets are aligned, by adding a constant to the $-\log(N_m/N_{tot})$ data in each set,  so that in the plot the last configuration of one data set coincides with the first configuration of the next data set.\footnote{The additive constants for the different data sets are in fact {\it required} to ensure continuity of the wave functional. Note that only the ratios $N_m/N_n$ correspond to ratios of the vacuum wavefunctional, as seen in \rf{relw}, and this ratio is insensitive to a constant added to all the $-\log(N_m/N_{tot})$ in any data set.  This is because the relative weights method does not determine the overall normalization of the Yang-Mills vacuum wavefunctional.  So there is always the freedom, with respect to a given data set,  to add an arbitrary overall constant to $-\log(N_m/N_{tot})$, and this freedom must be employed, in the case of many data sets, in order to satisfy the continuity of the wave functional.}
A single straight line, determined from the first data set, runs through all eight data sets.

\begin{figure}[t!]
\subfigure[~$\b=2.2$]  % caption for subfigure a
{   
 \label{b23}
 \includegraphics[scale=0.6]{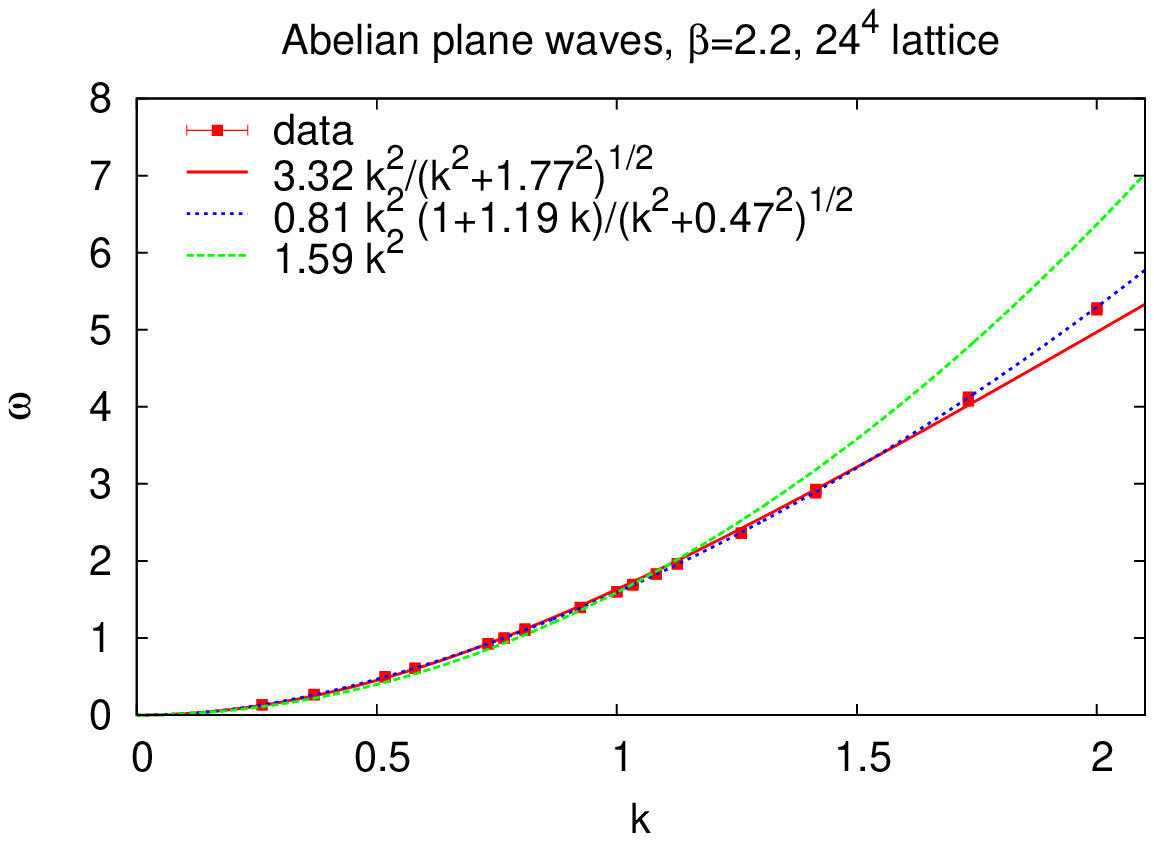}
}
\subfigure[~$\b=2.3$]  % caption for subfigure a
{   
 \label{b24}
 \includegraphics[scale=0.6]{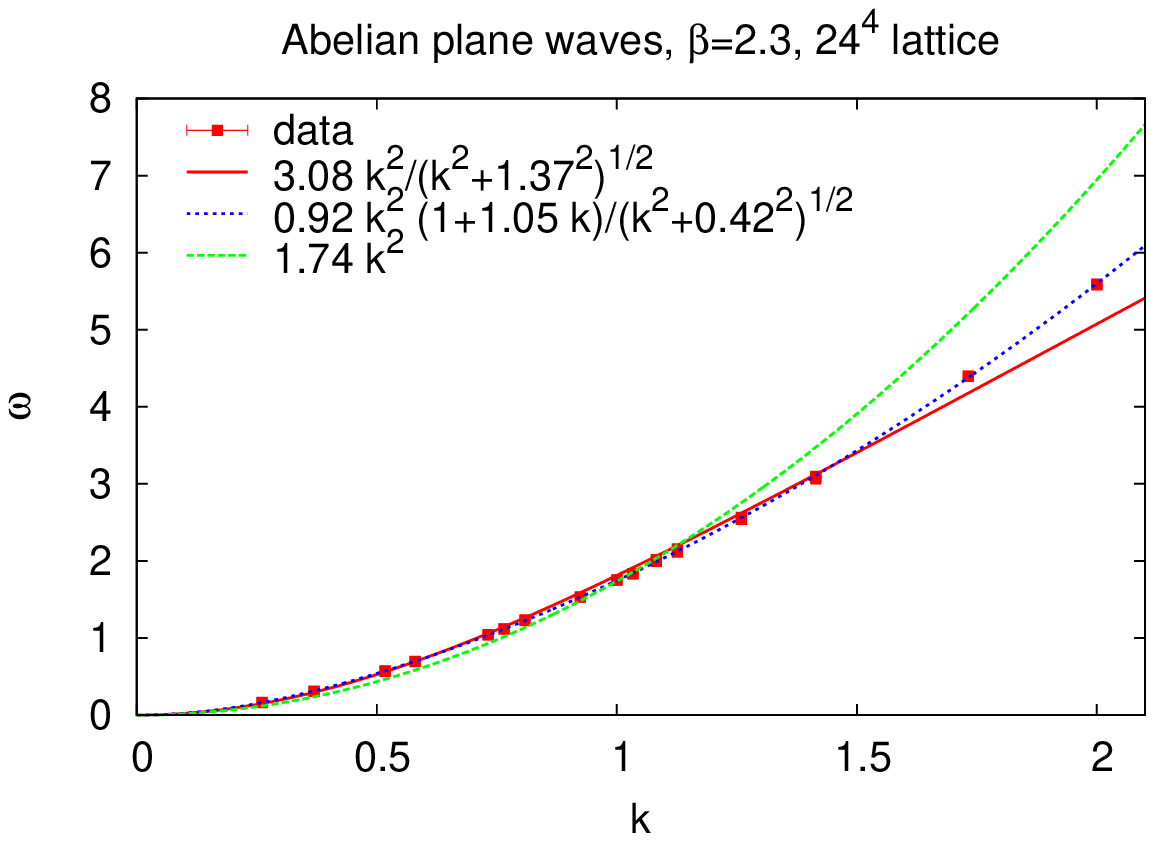}
}

\subfigure[~$\b=2.4$]  % caption for subfigure a
{   
 \label{b24}
 \includegraphics[scale=0.6]{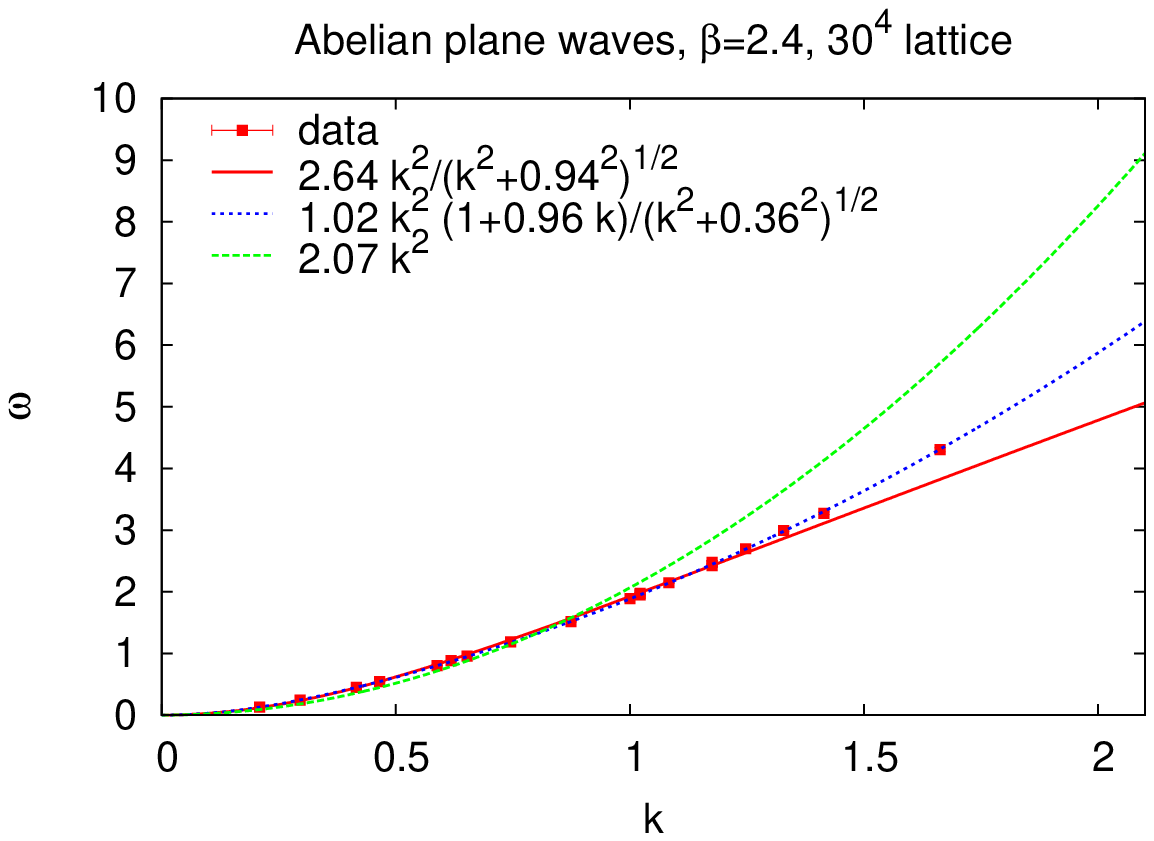}
}
\subfigure[~$\b=2.5$]  % caption for subfigure a
{   
 \label{b25}
 \includegraphics[scale=0.6]{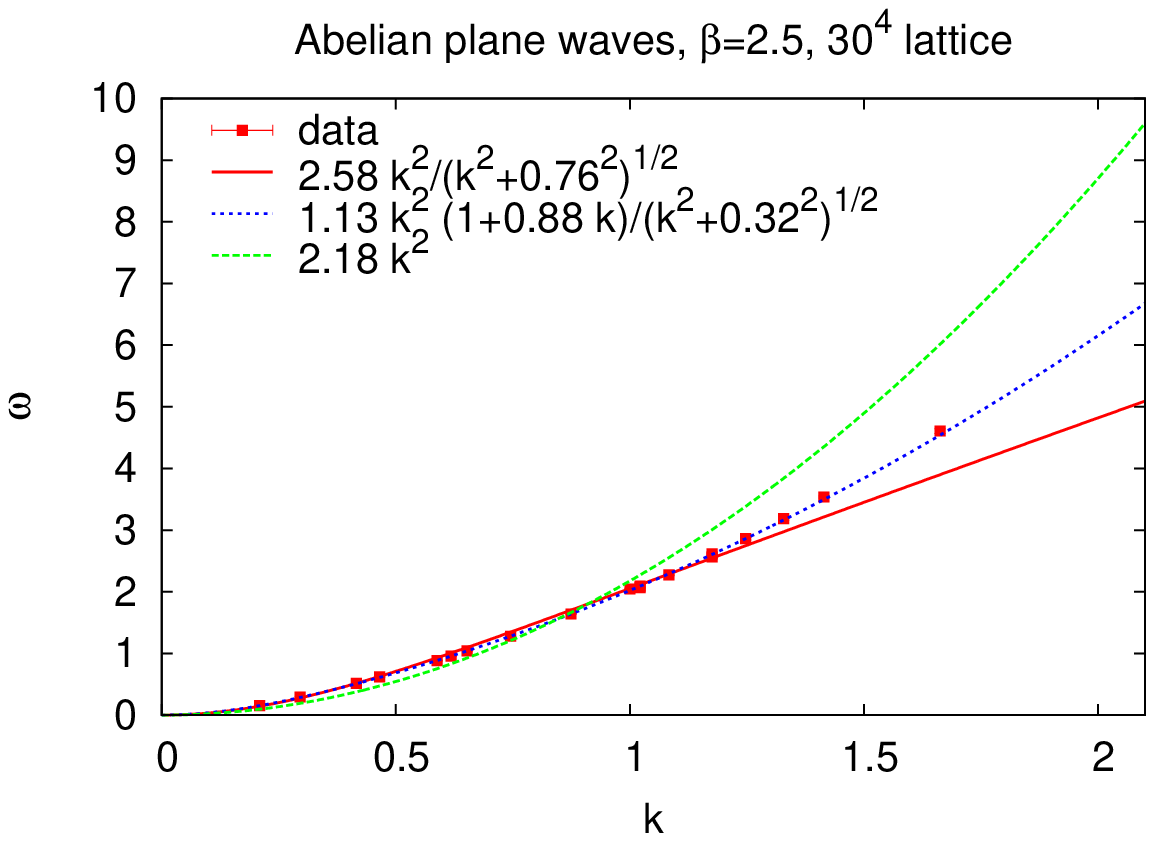}
}
\caption{$\o(k)$ vs.\ lattice momentum, here denoted $k$, for lattice couplings (a) $\b=2.2$, (b) $\b=2.3$, (c) $\b=2.4$
and (d) $\b=2.5$.  Data values are red dots.  Fits are shown for kernels corresponding to the original ansatz (eq.\ \rf{om1} with $d=0$, red curve), the generalized ansatz (eq.\ \rf{om1} with $d \ne 0$, blue curve), and the dimensional reduction limit (green curve).  }
\label{omega}
\end{figure}

Fig.\ \ref{omega}  displays our results for $\o(k_L)$ vs.\ $k_L$, at $\b = 2.2, 2.3, 2.4, 2.5$, versus
a fit to the form \rf{om1}, as well as a fit to the form $c k_L^2/\sqrt{k_L^2 + m^2}$, suggested by our original
ansatz \rf{PsiU}.  Also shown is the form $\o \propto k_L^2$ corresponding to the dimensional reduction limit.  The form \rf{om1} is clearly superior at the higher $k_L$ values. 

\newpage

\begin{figure}[t!]
\includegraphics[scale=1.0]{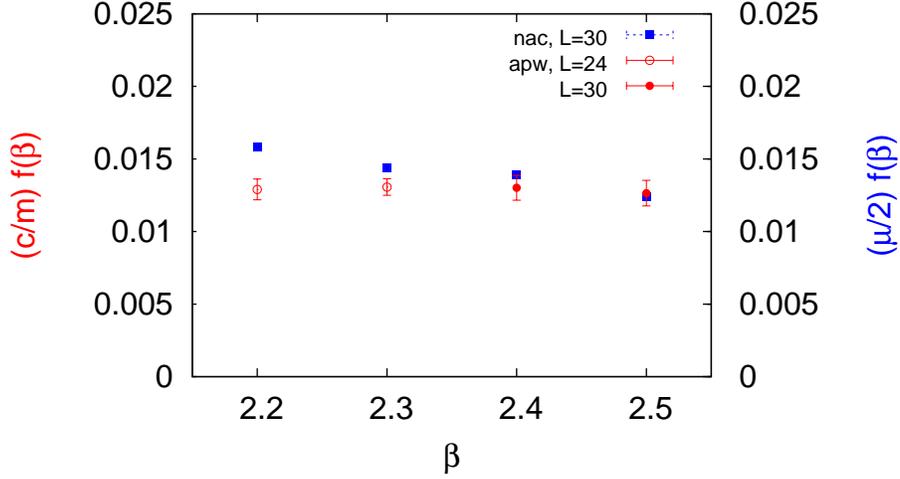}
\caption{Scaling test for $\m$ derived from non-abelian constant data (blue points), and $\m= 2c/m$ derived from the abelian plane wave data (red points) in the infrared limit. Both data sets are multiplied by the asymptotic freedom expression $f(\b)$ of eq.\ \rf{fb}.  If our wavefunctional is correct, these two rescaled data sets should coincide, and become $\b$-independent at sufficiently weak couplings.}
\label{NAC_vs_APW}
\end{figure}

   In the long-wavelength $k_L \ra 0$ limit, it is not hard to see that $R[U]$ in \rf{R0}, with either the  original or generalized version of $G(x-y)$, goes over to \rf{R1}, and likewise the dimensional  
reduction form \rf{R2}.   It is then of interest to compare $\m_{nac}$, derived from the non-abelian constant data, with the corresponding quantity $\m_{apw} = 2 c/m$, where $c$ and $m$ are extracted from the abelian plane wave data.  At sufficiently weak couplings, asymptotic freedom implies that $f(\b) 2c/m$ and $f(\b) \m$ should be constant with $\b$, and our wavefunctional implies that these quantities should equal one another. In Fig.\ \ref{NAC_vs_APW} we plot $f(\b) 2c/m$, obtained from the abelian plane wave data, and $f(\b) \m$, obtained from the non-abelian constant data, 
vs.\ lattice coupling $\b$.   The result is reasonably consistent with our expectations.   

\begin{figure}[htb]
\includegraphics[scale=0.8]{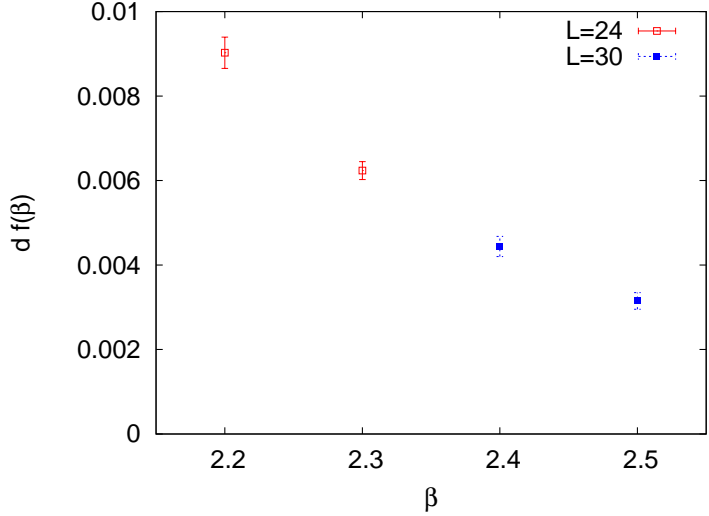}
\caption{$d f(\b)$ vs.\ $\b$.  The data indicates that the constant $d$ in physical units may vanish in the continuum limit.}
\label{d_vs_beta}
\end{figure}

    The generalization of the momentum kernel to finite $d$ can be accommodated by a revision of the gauge-invariant wavefunctional ansatz \rf{PsiU} to the form $\Psi_0^2 = \N \exp[-R[U]]$ with
\bea
R[U] &=& {c\over 4} \sum_x \sum_y \sum_{i<j} F^a_{ij}(x) G^{ab}_{xy} F^b_{ij}(y)  
\non \\ 
        &=& {c\over 8} \sum_x \sum_y  F^a_{ij}(x) \left( {1\over  \sqrt{-D^2 - \l_0 + m^2}} + d \sqrt{ -D^2 - \l_0 \over -D^2 - \l_0 + m^2} \right)^{ab}_{xy} F^b_{ij}(y) 
\label{modify}
\eea
However, for abelian configurations the momentum dependence of the generalized kernel is in complete disagreement with
that of the free theory at high momentum.  This means that the data seems to contradict the original motivation, which was to find a simple form interpolating between the free field and dimensional reduction expressions. 
On the other hand, inserting some powers of the lattice spacing
\bea
 R[U] &=& {c\over 8} \sum_x a^3 \sum_y a^3  \left({1\over a^2} F^a_{ij}(x) \right)  \left({1\over a^2}G^{ab}_{xy} \right)  \left({1\over a^2} F^b_{ij}(y) \right) 
\eea
we end up with the formal expression in the continuum limit
\bea
R[U] &=& {c\over 8} \int d^3 x \int d^3 y ~  \F^a_{ij}(x) {\cal G}^{ab}_{xy} \F^b_{ij}(y)  
\label{R}
\eea
where $\F_{ij} = F_{ij}/a^2,{\cal G}_{xy}=G_{xy}/a^2$ have the correct engineering dimensions in the continuum of 1/length${}^2$. Then we have
\bea
    {\cal G}^{ab}_{xy} = \left( {1\over  \sqrt{-D_{phys}^2 - \l_{phys,0} + m_{phys}^2}} + d_{phys} \sqrt{ -D_{phys}^2 - \l_{phys,0}
     \over-D_{phys}^2 - \l_{phys,0} + m_{phys}^2} \right)^{ab}_{xy}
\eea
where, with a lattice regularization, $D^2_{phys} = D^2/a^2, ~ \l_{phys} = \l/a^2, ~ d_{phys} = d a$.  If $d_{phys}$ is finite and nonzero in the continuum limit, then we would expect
\bea
          \lim_{\b \ra \infty}  d f(\b) = \mbox{finite and non-zero}
\eea
However, when we plot $d(\b) f(\b)$ we find the result shown in Fig.\ \ref{d_vs_beta}.  This data suggests that $d_{phys} = 0$ in the continuum limit, and it may be that the original form of the wavefunctional \rf{PsiU} is recovered in that limit.

\begin{figure}[t!]
\subfigure[]  % caption for subfigure a  
{
\includegraphics[scale=0.6]{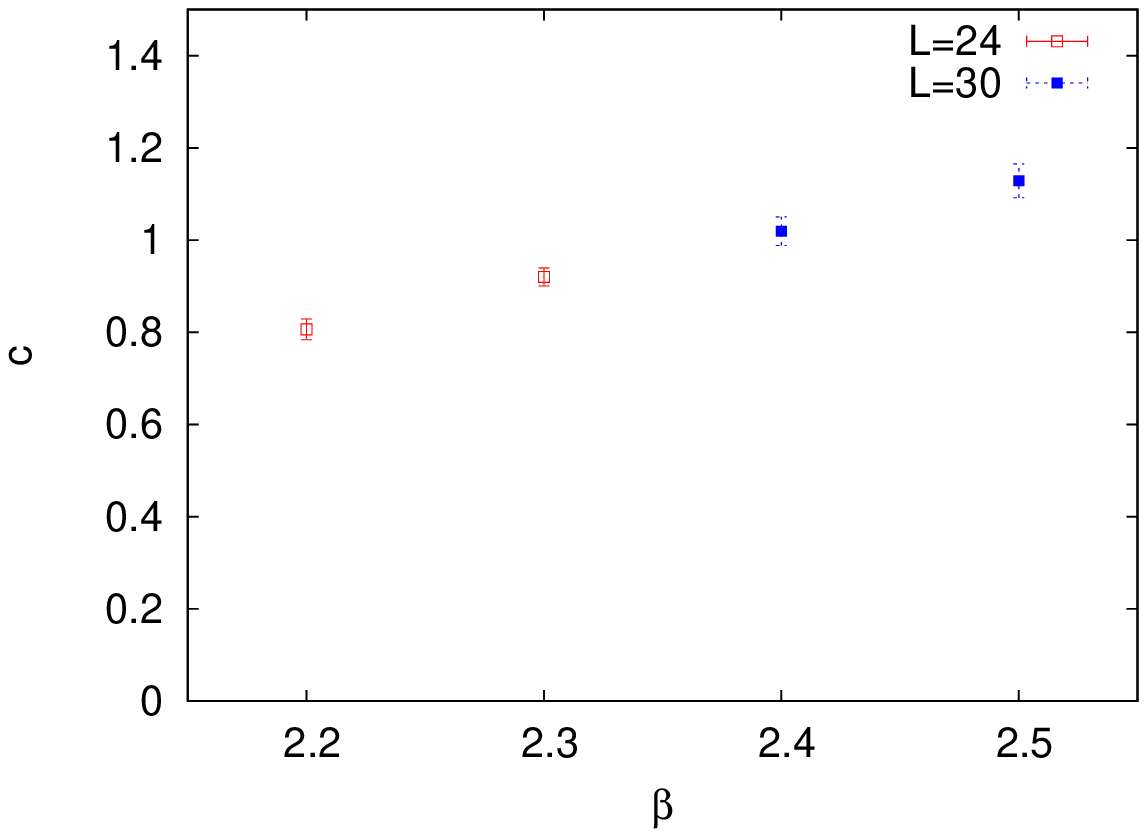}
}
\subfigure[]  % caption for subfigure a
{   
 \includegraphics[scale=0.6]{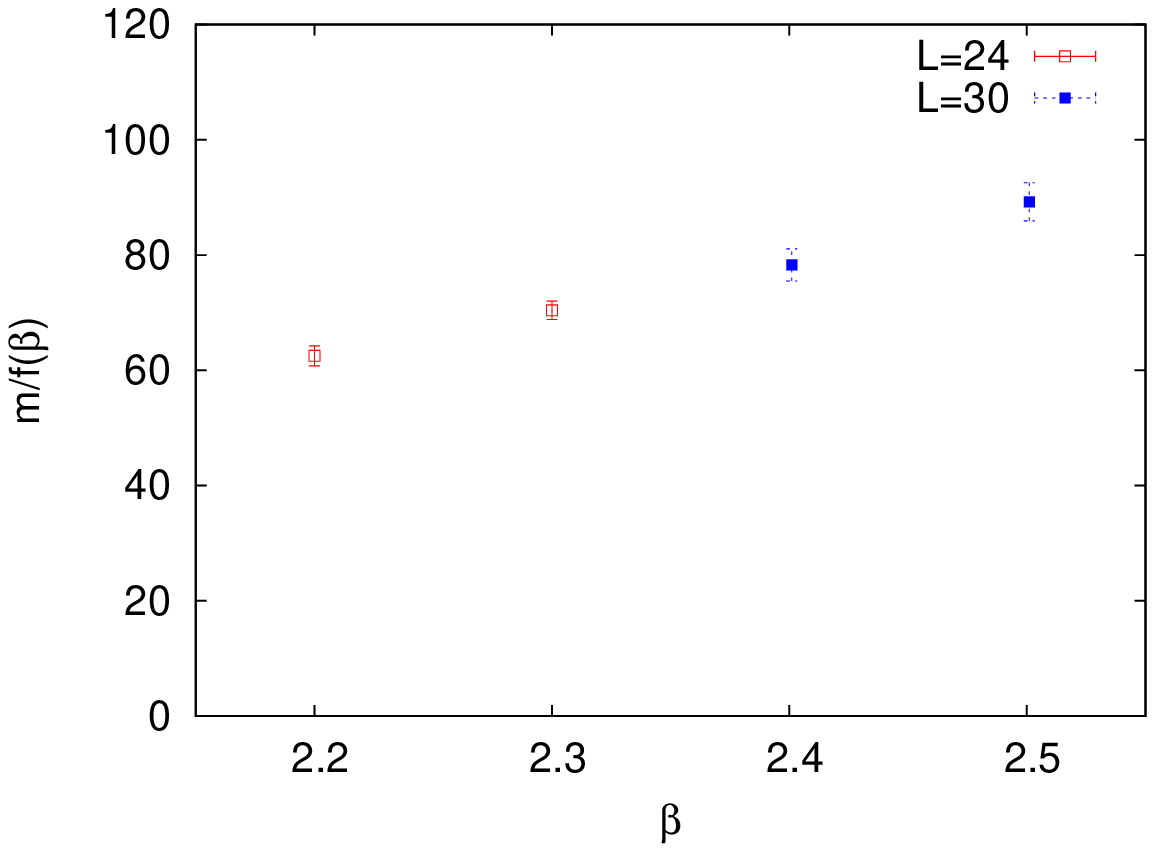}
}
\caption{Scaling tests for (a) parameter $c$ and (b) rescaled parameter $m/f(\b)$.  $\b$-independence of these quantities would indicate asymptotic scaling, which seems better for the ratio (Fig.\ \ref{NAC_vs_APW}) than for either quantity separately.}
\label{c_and_m}
\end{figure}

   It remains to check the variation with $\b$ of the parameters $c, m$, whose ratio $c/m$ has already been
seen, in Fig.\ \ref{NAC_vs_APW}, to scale in the correct way.  In Fig.\ \ref{c_and_m}  we plot $c$ vs.\ $\b$ and $m/f(\b)$ respectively.  The scaling is not as convincing for $c$ and $m$ separately, although the variation over the range of $\b=2.2-2.5$ is not so large, roughly on the order of 40\% and 50\% for $c$ and $m$ respectively, while the square root of the string tension in this range varies by about a factor of 2.5.

\section{Conclusions}

    With a modification (see \rf{modify}) which may disappear in the continuum limit, the conjectured vacuum wavefunctional
\rf{PsiU} on the lattice appears to be in harmony with vacuum amplitude data, obtained from the relative weights approach, for both non-abelian constant and abelian plane wave configurations.   For both non-abelian constant configurations and 
long-wavelength abelian plane wave configurations the vacuum wavefunctional reduces to the dimensional reduction form 
\rf{DR}, and the coefficient $\mu$, which amounts to the effective coupling of the action in one less dimension, is the same whether obtained from non-abelian constant configurations, or abelian plane wave configurations.

    One limitation of this work is that the configurations tested, non-abelian constant and abelian plane wave, are highly atypical. It would be preferable to apply the relative weights method to a set of small variations around a thermalized configuration.  We hope to carry out this generalization in a later study.
         
\acknowledgments{J.G.'s research is supported in part by the
U.S.\ Department of Energy under Grant No.\ DE-FG03-92ER40711.  \v{S}.O.'s research is supported by the Slovak Research and Development Agency under Contract No.\ APVV--0050--11, and by the Slovak Grant Agency for Science, Project VEGA No.\ 2/0072/13 (\v{S}.O.). In initial stages of this work, \v{S}.O.\ was also supported by ERDF OP R\&D, Project meta-QUTE ITMS 2624012002.}     

\bibliography{vac}

\begin{thebibliography}{1}

\bibitem{Greensite:1979yn}
J.~Greensite,
\newblock Nucl.Phys. {\bf B158}, 469 (1979).
%%CITATION = NUPHA,B158,469;%%

\bibitem{Halpern:1978ik}
 H. MatevosyanM.~Halpern,
\newblock Phys.Rev. {\bf D19}, 517 (1979).
%%CITATION = PHRVA,D19,517;%%

\bibitem{Greensite:1988rr}
J.~Greensite and J.~Iwasaki,
\newblock Phys.Lett. {\bf B223}, 207 (1989).
%%CITATION = PHLTA,B223,207;%%

\bibitem{Greensite:2007ij}
J.~Greensite and {\v S}.~Olejn\'{\i}k,
\newblock Phys.Rev. {\bf D77}, 065003 (2008), arXiv:0707.2860.
%%CITATION = ARXIV:0707.2860;%%

\bibitem{Samuel:1996bt}
S.~Samuel,
\newblock Phys.Rev. {\bf D55}, 4189 (1997), arXiv:hep-ph/9604405.
%%CITATION = HEP-PH/9604405;%%

\bibitem{Greensite:2011pj}
J.~Greensite, H.~Matevosyan, {\v S}.~Olejn\'{\i}k, M.~Quandt, H.~Reinhardt, and A.~Szczepaniak,
\newblock Phys.Rev. {\bf D83}, 114509 (2011), arXiv:1102.3941.
%%CITATION = ARXIV:1102.3941;%%

\bibitem{Karabali:1998yq}
D.~Karabali, C.-j. Kim, and V.~Nair,
\newblock Phys.Lett. {\bf B434}, 103 (1998), arXiv:hep-th/9804132.
%%CITATION = HEP-TH/9804132;%%

\end{thebibliography}
   
\end{document}